\documentclass[11pt]{article}

\usepackage{amsmath}
\usepackage{amssymb}
\newcommand{\Si}{\Sigma}
\newcommand{\tr}{{\rm TA}}

\newcommand{\db}{\bar{\partial}}
\newcommand{\bz}{\bar{z}}
\newcommand{\bw}{\bar{w}}
\newcommand{\Om}{\Omega}
\newcommand{\de}{\delta}
\newcommand{\al}{\alpha}
\newcommand{\te}{\theta}

\newcommand{\pa}{\partial}
\newcommand{\la}{\lambda}

\newcommand{\tq}{\tilde{q}}

\newcommand{\La}{\Lambda}

\newcommand{\cP}{{\cal P}}

\newcommand{\cA}{{\cal A}}
\newcommand{\bA}{{\bar A}}
\newcommand{\bP}{{\bf P}}
\newcommand{\bbP}{{\bf CP}}
\newcommand{\bC}{{\bf C}}

\newcommand{\ve}{\varepsilon}

\newcommand{\vf}{\varphi}

\newcommand{\ga}{\gamma}

\newcommand{\si}{\sigma}

\def\x {\stackrel {\textstyle \otimes}{,}}
\def\tr{{\rm tr}}
\newtheorem{teo}{Theorem}

\def\beq#1{\begin{equation}\label{#1}}
\def\eq{\end{equation}}
\newcommand{\beqn}[1]{\begin{eqnarray}\label{#1}}
\newcommand{\eqn}{\end{eqnarray}}

\def\f1#1{\frac{1}{#1}}

\begin{document}
\vspace{0.3in}

\title{Classical R-Matrices and \\[0.3cm]
the Feigin-Odesskii Algebra via\\[0.5cm]
Hamiltonian and Poisson
Reductions}
\author{H.W. Braden \thanks{School of Mathematics, University of Edinburgh, hwb@ed.ac.uk}
\and
 V.A. Dolgushev \thanks{Massachusetts Institute of Technology, Cambridge; 
Institute for Theoretical and Experimental
Physics, Moscow, vald@mit.edu}
\and
M.A. Olshanetsky \thanks{Institute for Theoretical and Experimental
Physics, Moscow, olshanet@heron.itep.ru}
\and
A.V. Zotov \thanks{Institute for Theoretical and Experimental
Physics, Moscow, zotov@heron.itep.ru}
}

\maketitle

\vspace{-10cm}
\begin{flushright}
 \begin{minipage}{1.2in}
 ITEP-TH-03/03\\
 EMPG-03-01
 \end{minipage}
\end{flushright}
\vspace{10cm}

\begin{abstract}
We present a formula for a classical $r$-matrix
of an integrable system obtained by Hamiltonian reduction 
of some free field theories using pure gauge symmetries.
The framework of the reduction
is restricted only by the assumption that
the respective gauge transformations are Lie group
ones. Our formula is in terms of Dirac brackets, and some
new observations on these brackets are made. We apply our method to derive
a classical $r$-matrix for the elliptic Calogero-Moser system
with spin starting from the Higgs bundle over an elliptic curve with
marked points.
In the paper we also derive a classical Feigin-Odesskii algebra
by a Poisson reduction of some modification of the Higgs
bundle over an elliptic curve.
This allows us to include integrable lattice models in a 
Hitchin type construction.

\end{abstract}

\section{Introduction}
A classical $r$-matrix structure is an important tool for investigating
integrable systems \cite{Sk79,STS,BV90}. It encodes the
Hamiltonian
structure of the Lax equation, provides the involution of integrals of motion,
and gives a natural framework for quantizing integrable systems in a
quantum group theoretic setting \cite{EV}.

The aim of this paper is severalfold.
First, we present a formula
for the classical $r$-matrices of integrable systems, derived
in the framework of Dirac's Hamiltonian reduction for systems
involving only gauge transformations.  In the process we shall derive
new results describing the Dirac brackets.  As an application of our
general formula we shall calculate a classical $r$-matrix for the 
elliptic Calogero-Moser system with spin.
Following the works \cite{Nikita,Olshanet,Dima}
we start from the Higgs bundle over an elliptic curve with a marked
point. This is a free theory with a trivial $r$-matrix. After the
Dirac procedure we come to the desired $r$-matrix.
Finally we derive the classical Feigin-Odesskii algebra \cite{FO}
in the similar fashion. Here the corresponding unreduced space is 
similar to the cotangent bundle of the centrally extended loop group 
$\hat{L}(GL_N)$.
The Poisson structure on this space is a particular example of
a general construction proposed in \cite{Po}.
In this way we derive the corresponding quadratic Poisson algebra
using the Hitchin approach \cite{Hit}. Thus we have managed to include 
integrable lattice models in the general Hitchin construction.

Our work is motivated by the papers
\cite{Babel,ArMed} where the authors use a gauge
invariant extension of Lax matrices to derive classical
$r$-matrices for the Toda
chain, and for the trigonometric and
elliptic (spin zero) Calogero-Moser systems.
In the first paper \cite{Babel}
the Hamiltonian reduction is considered on the cotangent
bundle to a finite dimensional Lie group,
while in the second paper the construction is
generalised to the case of the central extension
of the loop group.
In this context the paper \cite{Vengry} also
warrants mention. There
the authors consider a special case
of Poisson reduction on Poisson-Lie groupoids in
order to obtain new examples of the class
of dynamical $r$-matrices defined by
Etingof and Varchenko \cite{Etingof}.

In our paper we propose a very general framework for
Hamiltonian reduction by pure gauge symmetries
that allows one to derive a classical $r$-matrix.
This framework is based on rather general assumptions:
essentially that the first class constraints generate
the adjoint action of some Lie algebra on a (Lie algebra valued) function on
the unreduced space, which in turn reduces
to the Lax equation, and that the Poisson brackets between
these elements is already cast into $r$-matrix form.

The organisation of the paper is as follows.
In the second section we outline
the general framework of the
Hamiltonian reduction which allows us to
derive $r$-matrices for integrable
systems. In the third section we discuss the Dirac brackets and the $r$-matrices
more generally, relating each of these to generalized inverses. This sheds
some new light on Dirac's brackets. The fourth section consists of our
first example, where
we apply our method to derive the classical $r$-matrix
structure for the elliptic Calogero-Moser system
with spin \cite{Nikita,Dima} using
its \v{C}ech-like Hitchin description \cite{Hit,Olshanet}.
The fifth section
is devoted to the derivation of the
classical Feigin-Odesskii algebra \cite{FO} from some free field theory.
This theory is a modification of the Higgs bundle related to the
holomorphic $GL_N$-bundle of degree one over an
elliptic curve. Instead of a Higgs field that takes values in 
the endomorphisms of the bundle we consider the field of 
automorphisms of the bundle.
It is a Poisson manifold \cite{Po} and
we show that the elliptic Belavin-Drinfeld $r$-matrix \cite{BD}
is naturally obtained in the context of such a reduction.
In the concluding section we discuss a possible generalisation 
of our reduction technique to Hitchin systems on 
curves of higher genus. An Appendix is devoted to
the special elliptic functions we use.

Throughout the paper we use standard notations from 
quantum group theory and integrable systems to express the
Poisson brackets between entries of matrices.
Thus, for example, the $r$-matrix equation cast as
\begin{equation}
\{L\x L\}=
[r, L\otimes 1]-[{r\sp{T}},1\otimes L]
\label{rmatrix}
\end{equation}
becomes
\begin{equation}
\{L_1,L_2 \}=
[r_{12},L_1]-[r_{21},L_2].
\label{nota}
\end{equation}
If the Lax matrix takes values in some Lie algebra ${\mathfrak{g}}$ (or representation
of this),
the $r$-matrix takes values in ${\mathfrak{g}}\otimes{\mathfrak{g}} $ or its
corresponding representation. We may expand quantities in terms of a
basis of ${\mathfrak{g}}$ as follows.
Let  $t_\mu$ denote such a  basis
with $[t_\mu,t_\nu]=c_{\mu \nu}\sp\lambda\ t_\lambda$ defining the
structure constants of ${\mathfrak{g}}$. Suppose $\phi(t_\mu)=X_\mu$,  where
$\phi$ yields the desired
representation of the Lie algebra; we may take this
to be a faithful representation. With
$L= \sum_{\mu}L\sp\mu X_\mu $ the left-hand side of (\ref{rmatrix})
becomes
\begin{equation}
\{L\x L\}=\sum_{\mu,\nu} \{ L\sp\mu, L\sp\nu \} X_\mu\otimes X_\nu,
\end{equation}
while upon setting $r=r\sp{\mu\nu}X_\mu\otimes X_\nu$ and
${r\sp{T}}=r\sp{\nu\mu}X_\mu\otimes X_\nu$ the right-hand side
yields
\begin{align}
[r, L\otimes 1]-[{r\sp{T}},1\otimes L]&=
 r\sp{\mu\nu}L\sp\lambda
([X_\mu,X_\lambda]\otimes X_\nu-X_\nu\otimes[X_\mu,X_\lambda])\nonumber \\
&=( r\sp{\tau\nu}c_{\tau\lambda}\sp\mu L\sp\lambda -
    r\sp{\tau\mu}c_{\tau\lambda}\sp\nu L\sp\lambda) X_\mu\otimes X_\nu.
\label{rexplicit}
\end{align}

Concretely, in the standard basis of $gl_N$
$$(e_{ij})_{mn}=\de_{im}\de_{jn}$$
we have $\{X_\mu\}=\{e_{ij}\}$ and (\ref{nota}) takes the form
\begin{equation}
\{L_{ij},L_{kl}\}=
\sum_{m}(r_{imkl}L_{mj}-L_{im}r_{mjkl}-
r_{kmij}L_{ml}+L_{km}r_{mlij}),
\label{nota1}
\end{equation}
where
\begin{equation}
r=\sum_{i,j,k,l}r_{ijkl}\, e_{ij}\otimes e_{kl}.
\label{notta}
\end{equation}
Then $r_{21}$ denotes the function with
permuted indices $r_{21}(z)=r_{kl\,ij}(z)e_{ij}\otimes e_{kl}$.

At the outset let us record that
under a gauge transformation $r$-matrices transform as
\begin{equation}
\begin{split}
r^{G}(z,z')&=\La_1^{-1}\La_2^{-1}r(z,z')\La_1 \La_2
\label{rgauget}
+ \La_1^{-1}\La_2^{-1}\{l_2(z') ,\La_1\} \La_2   \\
&\qquad  + \frac12 [ \La_1^{-1}\La_2^{-1} \{\La_1 ,\La_2\},
\La_2^{-1} l_2(z') \La_2 ]\,.
\end{split}
\end{equation}
Also, given a Lax operator, $r$-matrices are far from being uniquely defined.
We will specify this ambiguity in due course.

\section{The Formal R-Matrix via Hamiltonian Reduction}
In this section we show that if a classical integrable system
is obtained by a Hamiltonian reduction involving purely gauge symmetries,
and so corresponding first class constraints, then
the system admits a canonical $r$-matrix.
The word ``formal'' in the title of the section stresses the fact that
in many interesting cases the reduction is performed in the context
of a field theory, and consequently for
such systems the formula for the $r$-matrix must be properly
defined.

Our approach makes three quite general assumptions regarding the Hamiltonian reduction
that allows us to calculate the $r$-matrix.
These are:
\begin{enumerate}

\item There exists an element $\Phi$, which upon reduction becomes
the Lax matrix, and this takes values in some Lie algebra $\mathfrak{g}$.

\item The gauge transformations
of $\Phi\in \mathfrak{g}$ can be represented as
\begin{equation}
\Phi \mapsto h^{-1}\Phi h\,,
\label{trans}
\end{equation}
where $h$ is assumed to take values in the Lie group $G$
corresponding to the Lie algebra $\mathfrak{g}$.

\item The  Poisson brackets between the
``entries'' of the element $\Phi$ in the unreduced space have been cast
into the $r$-matrix form before reduction.

\end{enumerate}

The first two assumptions are quite natural if we
wish to obtain an integrable system together
with its Lax representation. The second assumption implies that
the Poisson brackets between the first class constraints
$T_a$ generating the gauge symmetries, and $\Phi$
can be written in terms of the commutators
\begin{equation}
\{ T_a,\Phi \} = [ e_a, \Phi ],
\label{ono}
\end{equation}
where $\{e_a\}$ is a basis of the Lie algebra of gauge
transformations (which may be a subalgebra of $\mathfrak{g}$).
The third assumption is more elaborate.
It means that there exists a classical
$r$-matrix $r^0\in \mathfrak{g}\otimes \mathfrak{g}$, defining the
Poisson brackets in the total or unreduced space between the ``entries'' of
$\Phi$
\begin{equation}
\{\Phi_1,\Phi_2 \}=
[r^0_{12},\Phi_1]-[r^0_{21},\Phi_2].
\label{RO}
\end{equation}

Our first calculation below shows how the initial
$r$-matrix $r^0$ produces, under reduction, the desired classical
$r$-matrix of the integrable system. One may feel that we have
simply transferred the problem of constructing an $r$-matrix for
a reduced system to one of constructing an $r$-matrix for an unreduced
system. In practice constructing the latter $r$-matrix is often easier,
but we will deal with the construction of $r^0$ in the next section.

Together with some gauge fixing conditions $\chi^a=0$ the
first class constraints form a system of the second class
constraints $\{\si_{\al}\}=\{T_a,\chi\sp{b}\}$ \cite{Hen}. Using these
constraints we may define the Dirac bracket, which allows us to
calculate the reduced Poisson brackets in terms
of the Poisson brackets on the unreduced phase space. Thus, in order to
calculate the Poisson bracket between two
observable $f$, $k$ of the reduced phase space, we have first to
calculate the Dirac bracket between arbitrary
continuations $F$, $K$ of $f$ and $k$ in the unreduced space:
\begin{equation}
\{ F,K\}_{DB} =\{ F,K\}-
\{ F,\si_{\al}\}\, C^{\al\beta}\{ \si_{\beta},K\}.
\label{ona}
\end{equation}
Finally the reduced Poisson bracket
between $f$ and $k$ is obtained by restricting
expression (\ref{ona}) to the surface of
the second class constraints $\{\si_{\al}=0\}$.
Here $||C^{\al\beta}||$ is the matrix inverse to
$||C_{\al\beta}||=\{\si_{\al} ,\si_{\beta}\}$ and
$\{\,,\,\}$ is the Poisson bracket on the unreduced space.

Since under reduction the element $\Phi\in \mathfrak{g}$ becomes the Lax
matrix, the Poisson brackets between the ``entries'' of the Lax
matrix are just the on-shell Dirac brackets between the
respective  ``coordinates'' $\Phi$ on $\mathfrak{g}$.
Our third assumption automatically casts the first term in the
respective Dirac bracket (\ref{ona}) into the required form (\ref{RO}).
The  perhaps surprising thing is that the remaining
terms of the on-shell Dirac bracket (\ref{ona})
may also be cast into $r$-matrix form as a consequence of the second
assumption.
To see this we observe that (on-shell) $C$ takes the block form
\begin{equation}
C=\left(\begin{matrix} 0&\{T_a,\chi^b\} \\ \{\chi^a,T_b\}& \{\chi^a,\chi^b\}
  \end{matrix}\right)
 \equiv
\left(\begin{matrix} 0&P \\ -P\sp{T}& \{\chi^a,\chi^b\}
  \end{matrix}\right),
\qquad
P^b_a=\{T_a,\chi^b\},
\label{P}
\end{equation}
using the fact that $\{T_a,T_b\}=0$ as they are first class constraints.
Then $C\sp{-1}$ has the form
\begin{equation}
C\sp{-1}=\left(\begin{matrix} P\sp{T\, -1} \{\chi^a,\chi^b\} P\sp{-1}&-P\sp{T\,
-1}\\ P\sp{-1} &0 \end{matrix}\right)
 \equiv
\left(\begin{matrix} Q&-P\sp{T\,-1}\\ P\sp{-1} &0
  \end{matrix}\right),
\label{matC}
\end{equation}
where
\begin{equation}
Q^{ab}=(P^{-1})^a_c\{\chi^c,\chi^d \}  (P^{-1})^b_d .
\label{Q}
\end{equation}
Thus the only non-vanishing on-shell
entries of the matrix $||C^{\al\beta}||$ are
\begin{equation*}
C^{\chi^a\,T_b}=-C^{T_b\,\chi^a}=(P^{-1})^{b}_{ a},
\qquad\text{together with}\qquad
C^{T_a\, T_b}=Q\sp{ab}.
\label{2}
\end{equation*}
Using these the final term in $\{\Phi_1,\Phi_2\}_{DB}$ may be
written in $r$-matrix form as follows:
\begin{align*}
-\{  \Phi_1 ,\si_{\al}\}\, C^{\al\beta}\{ \si_{\beta}, \Phi_2 \}
&=-\{ \Phi_1 , \chi^a\} (P^{-1})^b_a \{T_b ,\Phi_2\}+
\{ \Phi_1 , T_a\} (P^{-1})^a_b \{\chi^b ,\Phi_2\}
\\ &\qquad -
\{ \Phi_1 , T_a\} Q^{ab} \{T_b ,\Phi_2\}\\
&
=-[(P^{-1})^a_b e_a\otimes \{\chi^b,\Phi\}, \Phi_1]+
[\{ \chi^a, \Phi \} (P^{-1})^b_a \otimes e_b ,\Phi_2 ]\\
&\qquad
+\frac{1}{2}[e_a \otimes Q^{ab} \{T_b ,\Phi\} ,\Phi_1]-
\frac{1}{2}[ \{\Phi, T_a\} Q^{ab} \otimes e_b ,\Phi_2]\,.
\end{align*}
Here we have made use of equation (\ref{ono}). Upon combining all of
the terms (\ref{RO}) for the on-shell Dirac bracket $\{\Phi_1,\Phi_2\}_{DB}$
we obtain the desired classical $r$-matrix.
\begin{teo} Under the three assumptions given we have
\begin{equation}
r=
(r^{0}-(P^{-1})^a_b e_a\otimes \{\chi^b,\Phi\}+
\frac12 Q^{ab} e_a\otimes \{T_b,\Phi\})\Big|_{\,on~shell}
\label{viaDirac}
\end{equation}
where $P$ is given in (\ref{P}) and $Q$ in (\ref{Q}).
\end{teo}

The terms modifying $r^0$ have the same general form as the infinitesimal
form of (\ref{rgauget}). We shall further explain this remark later 
in the paper.

In section four we apply the method outlined here to derive the 
classical $r$-matrix for the elliptic Calogero-Moser system with spin.
Before so doing however, let us return to the question of $r\sp0$,
and place our construction in a somewhat broader context.

\section{Dirac Brackets, Generalized Inverses and R-Matrices}
Our assumptions have {\it a priori} assumed the existence of an $r$-matrix
$r\sp0$ for the unreduced system.
In many examples such unreduced $r$-matrices are easy to construct: when the
unreduced matrices $\Phi$ depend only on half of the phase space variables, such as the
momenta, then for example $r^0=0$ is a possible solution.
Such however is not always the case, and in this section we shall discuss
the issue of the existence of $r^0$.
This will enable us both to place our Dirac Bracket construction
in a wider context and to elaborate on the remarks of \cite{B} concerning
this reduction procedure. For concreteness we shall phrase our discussion
in the language of finite dimensional matrices (as we did in the previous
section).

The construction of $r$-matrices is essentially an algebraic operation.
This algebraic nature was clarified in \cite{Br2} where a necessary and
sufficient condition was given for the existence of an $r$-matrix
based upon the fundamental Poisson brackets of the Lax matrix
(here $\Phi$). The novel part of this investigation was the use of
generalized inverses (and the construction thereof for generic elements
of the adjoint representation of a Lie algebra $\mathfrak g$).
We shall recall these notions in our present setting, relating them
to Dirac brackets. In so doing we will arrive at some new results pertaining
to Dirac brackets.

We begin with generalized inverses.
Let $A$ be an arbitrary matrix. (In particular
$A$ need neither be square nor invertible.)
A  matrix $G_A$ is said to be a generalized inverse of $A$ provided
\begin{equation}
A\,G_AA=A.
\label{geninverse1}
\end{equation}
Such a matrix always exists, yet need not be unique. We can further
require
\begin{equation}
G_A A\,G_A=G_A.
\label{geninverse2}
\end{equation}
Again such a generalized inverse always exists, yet need not be unique.
Observe that given  a $G_A$ satisfying (\ref{geninverse1}) and
(\ref{geninverse2}) we  have at hand
projection operators $P_1=G_AA$ and $P_2=A\,G_A$ which satisfy
\begin{equation}
AP_1=P_2A=A,\quad\quad P_1G_A=G_AP_2=G_A.
\label{proj}
\end{equation}
One may specify a unique generalized inverse (which always exists),
the  Moore-Penrose inverse, by additionally requiring
\begin{equation}
(A\,G_A)\sp\dagger=AG_A,\quad\quad (G_AA)\sp\dagger=G_AA .
\label{MP}
\end{equation}
Let us remark that the adjoint $()\sp\dagger$ here is defined with respect to
a given inner product, the  Moore-Penrose inverse
satisfying a norm-minimising condition.
Typically this is an hermitian inner product
and the adjoint is the hermitian conjugate.
We shall denote by $A\sp+$ the Moore-Penrose inverse of $A$.
(Accounts of generalized
inverses may be found in \cite{AG,Ca,Pr,RM}.)
Geometrically the Moore-Penrose inverse may be constructed by
orthogonally projecting onto the subspace on which $A$ has maximal
rank and inverting the resulting matrix.
We also record an alternative characterisation \cite{tevelev}.
Denote by $\tilde A$ and $\tilde G$ the matrices
$$\tilde A=\left(\begin{matrix} 0&A\\ 0&0\\ \end{matrix}\right),
\qquad \tilde G=\left(\begin{matrix} 0&0\\ G_A&0\\ \end{matrix}\right).
$$
Then
\begin{enumerate}
\item (\ref{geninverse1}) and (\ref{geninverse2}) $\Longleftrightarrow$
$\langle \tilde A, \tilde G, [\tilde A, \tilde G]\rangle$ form an
$sl_2$-triple.
\item Further $G_A=A\sp+$ $\Longleftrightarrow$
$[\tilde A, \tilde G]$ is Hermitian.
\end{enumerate}

Let us now relate Dirac brackets to generalized inverses.
We may suppose our (say unreduced) phase space has canonical
Poisson brackets
\begin{equation} \{ x\sp{i}, p_j \}=\de_{j}\sp{i}\label{canpb}
\end{equation}
so that
\begin{equation}
\{F,K\} =\left(\frac{\pa F}{\pa x},\frac{\pa F}{\pa p}\right)
\, J\, \left(\begin{matrix}  \dfrac{\pa K}{\pa x}\\ \\
\dfrac{\pa K}{\pa p}\end{matrix}\right), \qquad
J=\left(\begin{matrix} 0&1\\-1&0\end{matrix}\right).
\label{pbcoord}
\end{equation}
(We shall give a coordinate independent description in due course.)
In terms of the second class constraints $\sigma_\alpha$ we consider
the matrices
\begin{equation}
a_{\alpha}\sp{\,k}=\frac{\pa \sigma_\alpha}{\pa p_{k}},
\qquad
b_{\alpha k}=\frac{\pa \sigma_\alpha}{\pa x\sp{k}},
\qquad
C_{\alpha \beta}=\{\sigma_\alpha,\sigma_\beta\}.
\label{diracginv}
\end{equation}
In general we have a different number of constraints $c$ from the number
of coordinates, and the matrices $a$ and $b$ are not invertible.
Given that the Moore-Penrose inverse is constructed by an
orthogonal projection, and that Dirac's brackets give us a projection
onto the constraint surface, the following may not be surprising
(though it should be more widely known).
\begin{teo}  With the definitions (\ref{diracginv}) let $A=(b,a)$ and set
$G_A =\left(\begin{matrix}
a\sp{T}C\sp{-1} \\ -b\sp{T}C\sp{-1} \end{matrix}\right)$. Then
\begin{itemize}
\item[(i)] $$A\,G_AA=A,\qquad G_A A\,G_A=G_A\qquad
{\rm and}\qquad A\,G_A= {\rm Id}_{{\rm c}\times{\rm c}}=( A\,G_A)\sp\dagger.$$
\item[(ii)]  If $P_1=G_A\, A$ then $P_1\,J= J P_1\sp{T}$.
\item[(iii)] $$\{F,K\}_{DB}
=\left(\frac{\pa F}{\pa x},\frac{\pa F}{\pa p}\right)
\left(1-P_1\right)J \left(\begin{matrix}  \dfrac{\pa K}{\pa x}\\ \\
\dfrac{\pa K}{\pa p}\end{matrix}\right).$$
\item[(iv)] $P_1$ is self-adjoint with respect to any inner product of the form
$$\langle\langle {\bf u},{\bf v}\rangle\rangle=
{\bf u}\sp{T}P_1\sp{T}QP_1{\bf v}+
{\bf u}\sp{T}(1-P_1\sp{T})Q(1-P_1){\bf v}$$
where $Q$ is an invertible symmetric matrix. For such an inner product
$G_A=A\sp+$ is the Moore-Penrose inverse of $A$.
\end{itemize}
\end{teo}

The proof of these statements is straightforward. Observe that from the
basic Poisson bracket
\begin{equation}
C=b\, a\sp{T}- a\, b\sp{T}.\label{basicpb}
\end{equation}
we have that
$$
(b,a)\left(\begin{matrix}
a\sp{T}C\sp{-1} \\ -b\sp{T}C\sp{-1} \end{matrix}\right)=
(b\, a\sp{T}- a\, b\sp{T})C\sp{-1}= 1_{{\rm c}\times{\rm c}}.$$
Thus $P_2=A\, G_A=P_2\sp\dagger$ (for any choice of the adjoint).
It is then clear that
$G_A=\left(\begin{matrix} a\sp{T}C\sp{-1}\\ -b\sp{T}C\sp{-1}\end{matrix}\right)$
satisfies (\ref{geninverse1}) and (\ref{geninverse2}).
Direct calculation shows that
$$P_1=G_A\, A=\left(\begin{matrix} a\sp{T}C\sp{-1}b & a\sp{T}C\sp{-1}a\\
-b\sp{T}C\sp{-1}b & -b\sp{T}C\sp{-1}a \end{matrix}\right),$$
whence $P_1\,J= J P_1\sp{T}$, and $(iii)$ follows from (\ref{ona}).
We will have established the theorem upon showing $P_1$ is self-adjoint.
It is now that the issue of inner product confronts us. We shall discuss this
more generally after proving the final assertion of the theorem .
Now using the specified inner product and that $P_1$, $P_1\sp{T}$ are
projectors we have that
$$
\langle\langle P_1{\bf u},{\bf v}\rangle\rangle=
{\bf u}\sp{T}P_1\sp{T}\cdot P_1\sp{T}QP_1{\bf v}=
{\bf u}\sp{T}P_1\sp{T}QP_1{\bf v}=
{\bf u}\sp{T}P_1\sp{T}QP_1\cdot P_1{\bf v}=
\langle\langle {\bf u},P_1{\bf v}\rangle\rangle.
$$
Then upon comparing with the definition of the adjoint,
$\langle\langle P_1{\bf u},{\bf v}\rangle\rangle=
\langle\langle {\bf u},P_1\sp\dagger{\bf v}\rangle\rangle$,
we see that $P_1=P_1\sp\dagger$ and so is self-adjoint. Therefore $G_A=A\sp+$ is
the Moore-Penrose inverse for such an inner product. $\square$

Let us now describe this theorem in a more geometric manner. Denote
the unreduced phase space by $(P,\omega)$. Given a function $F$ on $P$ we have
a corresponding vector field $X_F$ given via
$$dF=i_{X_F}\,\omega=\omega(X_F,\ ).$$
The Poisson bracket of two functions is then
$$\{F,K\}=\omega(X_F,X_K)=dF(X_K).$$
Thus with $\omega=\sum dq\sp{j}\wedge dp_j$ we have
$$X_F= \frac{\partial F}{\partial p_j}\frac{\partial}{\partial x\sp{j}}
-\frac{\partial F}{\partial x\sp{j}}\frac{\partial}{\partial p_j}
$$
and the Poisson brackets (\ref{pbcoord}). Our constraints describe
a reduced phase space $V=\cap_{\alpha=1}\sp{c}\{\sigma_\alpha =0\}\subset P$.
Important for us is that $V$ is not an arbitrary submanifold but a symplectic
submanifold, and the inclusion
$i:V\hookrightarrow P$ gives the symplectic form $\omega_V=i\sp* \omega$ on
$V$. Because $\omega_V$ is nondegenerate,
$T_z V\cap (T_zV)\sp\omega=0$, where
$(T_zV)\sp\omega=\{u\in T_zP\,|\, \omega(u,v)=0\ \forall v\in V\}$.
Then
$$T_zP=T_z V\oplus (T_zV)\sp\omega$$
and we have a projection $\pi:T P\rightarrow T V$. One finds that
$$\pi( X_F)=X_F- \{F,\sigma_\alpha\}\, C^{\alpha \beta}\, X_{ \sigma_\beta}
=X_F-X_{ \sigma_\beta}\, C^{\beta\alpha}\, d\sigma_\alpha(X_F)=
[1-X_{ \sigma_\beta}\, C^{\beta\alpha}\, d\sigma_\alpha](X_F)
$$
and the Dirac brackets are
$$
\{F,K\}_{DB}=\{F|_V, K|_V\}=\omega(\pi( X_F),\pi( X_K))= dF\big(\pi( X_K)\big)
=dF\big([1-X_{ \sigma_\beta}\, C^{\beta\alpha}\, d\sigma_\alpha](X_K)\big)
$$
(Observe that $\pi\sp2=\pi$ follows from $d\sigma_\alpha(X_{ \sigma_\beta})
=C_{\alpha \beta}$. Such a projection operator will exist even for a
Poisson manifold $P$ provided $C_{\alpha \beta}$ is invertible: in this case
we have $T_zP= {\rm ker}\,\pi\oplus {\rm Im }\,\pi$ and ${\rm Im }\,\pi$ is
again Poisson.)
Comparison with the corresponding coordinate expression given by $(iii)$
of the theorem then shows that
$$\pi = 1-P_1.$$
The projection operator $P_1$ is then $P_1:T P\rightarrow (T_zV)\sp\omega$,
which is spanned by the vector fields $\{X_{ \sigma_\alpha}\}$. Everything
thus far has taken place within the realm of symplectic geometry:
$P_1$ is a symplectic projector and is self-adjoint with respect to the
nondegenerate bilinear form $\omega$,
$$\omega(P_1( X_F), X_K)= \omega(P_1( X_F), P_1(X_K))=
\omega( X_F, P_1(X_K))=\omega( X_F, P_1\sp\dagger(X_K)).
$$
(This is the content of $(ii)$ of the theorem with $\omega({\bf u},{\bf v})
={\bf u}\sp{T} J {\bf v}$.)
To talk of orthogonal projection we need an inner product. While the
kinetic energy of a natural Hamiltonian system can provide this 
an inner product is an additional ingredient. However, as $(iv)$ of the theorem
shows, there are a large class of inner products for which $\pi$ becomes
an orthogonal projection and we have
\begin{equation}T_zP=T_z V\oplus (T_zV)\sp\perp.\label{vsdecomp}\end{equation}
In practice one often further restricts attention to compatible inner products
for which we have
$$\langle\langle  {\bf u},{\bf v}\rangle\rangle= \omega(J{\bf u},{\bf v}),
\qquad  \omega({\bf u},{\bf v})=\omega(J{\bf u},J{\bf v}),
\qquad  J\sp2=-{\rm Id}
$$
and respecting the vector space decomposition (\ref{vsdecomp}).

We remark that generalized inverses have been discussed in various
connections with singular Lagrangian systems. Broadbridge and Petersen
\cite{BP} refine an observation of Duffin \cite{Du}
that a generalized inverse may be used to go from a singular
Lagrangian system, such as those arising
when constraints are implemented by Lagrange multipliers, and a
corresponding Hamiltonian system. The reduced Hamiltonians appearing correspond
to those of de Leeuw {\it at al} \cite{LPP}. (These works place various
restrictions on the nature of the constraints.) Because the Moore-Penrose
inverse may be calculated very efficiently their use in solving constrained
dynamical systems is important. Kalaba and Udwadia formulate a large class
of constrained Lagrangian systems as a quadratic programming problem, and
use the Moore-Penrose inverse to solve these \cite{KU}.

Finally, just as the Dirac bracket can be expressed in terms of generalized
inverses, so too can the solution of the $r$-matrix equations (\ref{nota}) or
(\ref{RO}). From (\ref{rexplicit}) the $r$-matrix equation takes the
explicit form
$$
b=a\sp{T} r-r\sp{T} a
$$
where we have set
$ a\sp{\mu\nu}= c_{\mu\lambda}\sp\nu L\sp\lambda \equiv-ad(L)\sp\nu _\mu$
and
$ b\sp{\mu\nu}=\{ L\sp\mu, L\sp\nu \}$.
The solutions of this equation have been studied \cite{Br3}.
\begin{teo} Let $g$ be a generalized inverse of $a=-ad(L)$ satisfying
(\ref{geninverse1}) and (\ref{geninverse2}). Then, with $P_1=g\,a$ and $P_2=a\,g$,
the $r$-matrix equation
(\ref{rmatrix}) has solutions if and only if
\begin{equation}
 (1-P_1\sp{T})\,b\,(1-P_1)=0,
\label{consistency}
\end{equation}
in which case the general solution is
\begin{equation}
r={1\over2} g\sp{T} b P_1+g\sp{T} b(1-P_1) +(1-P_2\sp{T})Y+ (P_2\sp{T}ZP_2)a
\label{gensoln}
\end{equation}
where $Y$ is arbitrary and $Z$ is only constrained by the requirement that
$P_2\sp{T}ZP_2$ be symmetric.
\end{teo}
The theorem then constructs the $r$-matrix in terms of a generalized
inverse to $ad L$ and describes the ambiguity of the solution.
Although the general solution appears to depend on the generalized
inverse, the work cited shows that changing the generalized inverse only
changes the solution by such terms. Further, the generalized
inverse of a generic $ad L$ may be constructed.
Thus the existence of an $r$-matrix has been reduced to the
single consistency equation (\ref{consistency}) and the construction of a generalized
inverse to $ad(L)$.

Having now addressed the issue of $r\sp0$, and indeed the general
construction of $r$-matrices, we will return to the particular construction
given by the Dirac reduction procedure of the previous section.
We shall illustrate this in turn by two examples.

\section{Example 1: Derivation of the R-Matrix for the Elliptic Ca\-lo\-ge\-ro-Mo\-ser
System with Spin}
In this section we show that a classical $r$-matrix for the
elliptic Calogero-Moser system with spin 
can easily be derived with the help
of Dirac brackets. One surprise (for us) is that in
calculating the $r$-matrix we do not make a use of the
identities for elliptic functions that typically underly
these integrable systems.

In deriving the $r$-matrix for the elliptic Calogero-Moser 
system with spin we will divide the 
Hamiltonian reduction procedure into two stages.
In the first stage we omit gauge transformations related
to the parabolic subgroup. This yields the formula for 
the classical $r$-matrix of the
spin Calogero-Moser system with the maximal 
spin sector.

After completing the first stage of the reduction
we are left with residual action of the parabolic 
subgroup on a finite dimensional phase space. 
Here it is possible to derive the desired classical $r$-matrix
using any concrete fixing of the residual gauge symmetry.
An alternative method also exists, making use of
a gauge invariant extension of the Lax matrix. We will adopt the
latter approach to show that upon further reduction the 
$r$-matrix obtained in the first stage becomes a classical 
$r$-matrix for the elliptic spin Calogero-Moser 
system.
 
The Hamiltonian reduction that leads to the elliptic Calogero-Moser system
with spin was originally presented in the
paper \cite{Nikita}, where the phase space of the
system was realised as the cotangent bundle to the
moduli space of topologically trivial holomorphic bundles
(the Higgs bundles) over the torus with a marked point. 
We are going to employ this
construction in deriving the respective classical $r$-matrix.

For our purposes we mostly follow the paper \cite{Olshanet}
and use a  \v{C}ech like description of the
moduli space.
We will realise the torus $\Si_\tau$ as a quotient of ${\bf C}/q{\bf Z}$
 with $q=e^{2\pi i\tau},~{\rm Im}\tau>0$. We choose as the fundamental domain the annulus
$Ann_\tau=\{|q^{1/2}|<|z|<|q^{-1/2}|\}$.

We define a holomorphic vector bundle $E_N$ 
over $\Si_\tau$ by the matrix transition function $g(z)\in GL_N(\bC)$ that is
holomorphic in a neighbourhood of the contour $\ga=\{|z|=|q^{1/2}|\}$.
In this way $g(z)$ represents a \v{C}ech cocycle.
The set of these fields we denote as ${\cal L}=\{g\}$.

Let ${\mathcal G}$ be the gauge group of holomorphic maps $f(z)$ of the annulus $Ann_\tau$ to
$GL_N(\bC)$ such that
\beq{gt}
f(z)\Big|_{z=1}=I,
\end{equation}
where $I$ is the identity matrix. It acts on ${\cal L}$ as
\beq{ga}
g(z)\mapsto f(z)g(z)f^{-1}(q^{-1}z)\,.
\end{equation}
Observe that the Lie algebra Lie$({\mathcal G})$ consists of matrix-valued 
holomorphic functions $\ve(z)\,:\,Ann_{\tau}\mapsto\,gl_N(\bC)$ vanishing
at the point $z=1$
$$
\ve(z)\Big|_{z=1}=0.
$$

The quotient space under the gauge action (\ref{ga})
\beq{a18}
{\cal M}= {\cal L}/{\cal G}
\end{equation}
is the moduli space of the holomorphic bundles over
$\Si_\tau$ with the marked point $z=1$. 
Note that upon reduction with respect to 
adjoint action of a parabolic subgroup 
$\cP\subset SL_N(\bC)$ the space (\ref{a18}) 
becomes the moduli space of the holomorphic bundles over
$\Si_\tau$ with a quasi-parabolic structure at $z=1$.

The cotangent bundle $T^*{\cal L}$ is called the Higgs
bundle.
The dual field (the Higgs field) is a  one-form
$$\eta=\eta(z)\frac{dz}{z}$$ 
taking values in $End~(E_N)=$ Lie algebra $gl_N$.
The field $\eta(z)$  is holomorphic in a neighbourhood of $\ga_1$.

The bundle $ T^*{\cal L}$  is the non-reduced phase space with
the symplectic form
\beq{sf}
\Omega=\f1{2\pi i}\oint_{\ga_1}\tr\delta (g^{-1}(z)\eta)\wedge\delta g(z)\,.
\end{equation}
We lift the gauge group action (\ref{ga}) on $T^*{\cal L}$ as
\begin{equation}
\eta(z)\mapsto f(z)\eta(z)f^{-1}(z)\,.
\label{gaugeegm}
\end{equation}
The symplectic form (\ref{sf}) is invariant with respect to
this gauge action. In this way it produces the moment map
\beq{Mommap}
\mu~:~T^*{\cal L}\mapsto Lie^*({\mathcal G})\,,
\end{equation}
which is defined as 
\beq{Mommap1}
\mu[\eta,g](\ve)=\oint_{\ga_1} \tr
(g^{-1}(z)\,\eta\, g(z)\, \ve(q^{-1}z) - \eta\,\ve(z)),\qquad 
\ve \in Lie({\mathcal G}).
\end{equation}

The zero level of the moment map (\ref{Mommap1}) is the 
surface in $T^*{\cal L}$ where $\eta(z)$ can be extended
meromorphically from the boundary $(\ga_1,~\ga_2)$ inside the annulus
with at most single pole of the first order at the 
point $z=1$ and such that for $|z|=|q^{1/2}|$ 
\beq{qqqperiod}
\eta(q^{-1}z)= g^{-1}(z)\eta(z)g(z).
\end{equation}

While a generic topologically trivial bundle over a
torus can be decomposed into a direct sum of
line bundles \cite{Atiyah}, a generic holomorphic 
bundle with a marked point can differ from a
direct sum of line bundles by the conjugation 
of a constant matrix $\al\in SL_N(\bC)$.
In other words, using the gauge group action (\ref{ga})
one can transform a generic field 
$g(z)$ to a constant matrix of the following form 
\beq{0.1}
h=\al 
\left(\begin{array}{cccc}
e^{-2\pi i u_1} & 0 &\ldots & 0 \\
0 & e^{-2\pi i u_2}& \ldots & 0 \\
\ldots & \ldots & \ldots & \ldots \\
0 & 0 & \ldots & e^{-2\pi i u_N}
\end{array}\right)\al^{-1} \,.
\end{equation}
In particular, this means that 
\begin{equation}
\pa_z g_{ii}(z)=0
\label{GF111}
\end{equation}
is a proper gauge condition for our system.   

It is easy to see that the entries of the matrix  
$\al$ are defined by $h$
up to the scaling transformations

\begin{equation}
\al_{ij}\mapsto \al_{ij}\la_j, \qquad \la_j\neq 0.  
\label{scale}
\end{equation}   
while $u_i$ and $u'_i=u_i+n_1 + n_2\tau\,,~n_1,n_2\in {\bf Z}$ 
gives the same holomorphic bundle over $\Si_{\tau}$.   
Thus, an open dense set of the 
moduli space ${\cal M}$ (\ref{a18}) can be parameterised by 
the points of the set\footnote{In fact we slightly abuse the parameterisation
omitting the quotient with respect to the symmetric group of 
permutations $S_N$, which plays an analogous role as the Weyl group 
for the rational Calogero-Moser system.} 
\begin{equation}
(J\times \bbP^{N-1})^N,  
\label{eq1}
\end{equation}
where $J$ is the Jacobian of the curve 
$\Si_{\tau}$.
In the general context of holomorphic vector bundles over Riemann surfaces
this is known as the Tyurin parameterisation \cite{AT}. This parameterisation
has found recent application in a general description of Hitchin systems 
and their $r$-matrices \cite{ER1}, \cite{IK}, \cite{VAD}.

Expanding the functions $g(z)$ and $\eta(z)$ in a neighbourhood
of the contour $\ga$,
\begin{equation}
\eta_{ij}(z)=\sum_{a\in {\bf Z}}\eta^a_{ij} z^a\,, \qquad
g_{ij}(z)=\sum_{a\in {\bf Z}} g^a_{ij} z^a
\label{Loran}
\end{equation}
we readily find the respective Poisson brackets to be
\begin{equation}
\{\eta^a_{ij} , \eta^b_{kl}\}=-\eta^{a+b}_{il}\de_{kj} +
\eta^{a+b}_{kj}\de_{il},
\qquad
\{ g^a_{ij}, \eta^b_{kl}\}= g^{a+b}_{kj}\de_{il},
\qquad
\{ g^a_{ij}, g^b_{kl}\}=0.
\label{PB}
\end{equation}
In these terms the expansion coefficients of the constraint
$\mu[\eta,g]=0$ (\ref{Mommap1}) are of the
form
\begin{equation}
M^a_{ij}=\sum_{b,c,k,l}
(g^{-1})^{b-c}_{ik}\eta^{c}_{kl}g^{a-b}_{lj}q^{a}-
\sum_{b,c,k,l}
(g^{-1})^{b-c}_{ik}\eta^{c}_{kl}g^{-b}_{lj}
-\eta^a_{ij}+\eta^0_{ij}\,,\qquad a\neq 0, 
\label{MM}
\end{equation}
and the gauge fixing conditions (\ref{GF111})
can be rewritten as follows
\begin{equation}
g^a_{ij}=0, \qquad a\neq 0.
\label{GF}
\end{equation}

Our second class constraints are then
\begin{equation}
M^a_{ij}=0~(a\neq 0),\ \ \
g^a_{ij}=0~(a\neq 0).\ \ \
\label{Constr}
\end{equation}
The field $g(z)$ being restricted to
the constraint surface (\ref{Constr}) becomes 
the constant matrix $g(z)=h$ (\ref{0.1}) while the field 
$\eta$ yields the Lax matrix 
of the elliptic Calogero-Moser system
with maximal spin sector,  
\begin{equation}
\eta(z)=l(z)=\al\, \tilde l(z)\, \al^{-1} 
\label{Lax}
\end{equation}
with
$$
\tilde l_{ii}(z)=-\frac{v_i}{2\pi i}, \qquad
\tilde l_{ij}(z)=-\sum_{k=1}^N\beta_{ik}\al_{kj}\,\phi(z,u_{ij}),~i\neq j\,, 
\qquad u_{ij}=u_i-u_j\,.
$$
Here $u_i$, $\al_{ij}$, $v_j$ and $\beta_{ij}$ are
coordinates on the reduced phase space 
$T^*{\cal M}\approx T^*(J\times \bbP^{N-1})^N$.
Because $\al_{ij}$ (for each fixed $j=1,\,\ldots\,,\,N$)
are homogeneous coordinates on the respective projective spaces
$\bbP^{N-1}$, the symplectic form on $T^*{\cal M}$
is obtained from the form
\begin{equation}
\Om_1= \sum_{i=1}^N dv_i \wedge du_i + \sum_{i,j=1}^N
d\beta_{ij}\wedge d\al_{ji}  
\label{form-red}
\end{equation}
by symplectic reduction on the first class constraint surface
(arising from the generators of (\ref{scale}))
\begin{equation}
\sum_{k=1}^N \beta_{ik}\al_{ki}=0\,.
\label{gen-scale}
\end{equation} 
The function $\phi(z,u)$ in (\ref{Lax}) can be represented
in the following form
\begin{equation}
\phi(z,u)=\sum_{a\in{\bf Z}} \frac{z^a  }
{q^a e^{2\pi i  u }-1}, \quad |q|<|z|<1\,, \quad u\neq a+b\tau,~
a,b\in{\bf Z}.
\label{Ser1'}
\end{equation}

We note that the Lax matrix of the Hitchin system with 
a quasi-parabolic structure at the marked point $z=1$ is 
obtained from (\ref{Lax}) by reduction with respect to the adjoint action of 
the corresponding parabolic subgroup $\cP\subset SL_N(\bC)$\,.

In order to construct the $r$-matrix using our theorem we
must determine the Dirac bracket. First we calculate
the matrix of the Poisson brackets between
the constraints (\ref{Constr}) and then determine the inverse
matrix.
The non-vanishing on-shell brackets are found to be
\begin{equation}
\{g^a_{ij}\, ,M^b_{kl} \}=\de_{a+b,0}\,
(q^{-a}h_{il}\de_{kj}-h_{kj}\de_{il}).
\label{MPB}
\end{equation}
To present the inverse matrix we adopt the
following convention for the entries of the
inverse matrix: $(C^{Mg})^{ij~kl}_{a~b}$
stands for the entry with $M_{ij}^a$ being the first index and
$g_{kl}^b$ being the second. Then the
non-vanishing entries of the
inverse matrix can be written in the following form
\begin{equation}
(C^{Mg})^{ij~kl}_{a~b}= -
(C^{gM})^{kl~ij}_{b~a}=\sum_{m,n=1}^N
\frac{\de_{a+b,0}\,\al_{lm}
(\al^{-1})_{mi}\al_{jn}(\al^{-1})_{nk}}{q^a\,e^{-2\pi iu_n}-e^{-2\pi i u_m}}\,.
\label{InV}
\end{equation}

In deriving the $r$-matrix via the Dirac bracket we
will exploit the following simple property
\begin{equation}
\{M^a_{ij},\eta^b\}=[e_{ji},\eta^{a+b}]-[e_{ji},\eta^b],
\label{Group}
\end{equation}
which is just the counterpart of
the general relation (\ref{ono}).

Now the expression for the Dirac bracket (\ref{ona}) between the
expansion coefficients of the field $\eta(z)$ consists
of two parts. The first part is just the
initial Poisson bracket in the extended phase space. This may
be rewritten as
\begin{equation}
\{\eta^a_1,\eta^b_2\}=\frac12\sum_{i,j}
[e_{ij}\otimes e_{ji},\eta^{a+b}_1]-\frac12\sum_{i,j}
[e_{ij}\otimes e_{ji},\eta^{a+b}_2].
\label{PBLL}
\end{equation}
The second part reflects the reduction. With
$\{\eta^a_1,\eta^b_2\}_{DB}=\{\eta^a_1,\eta^b_2\}+D^{ab}$
the on-shell expression for the second part is
\begin{align}
D^{ab}= &\sum_{m,n=1}^N \frac{\al_{im}(\al^{-1})_{ml}\al_{kn}(\al^{-1})_{nj}}
{q^b\, e^{-2\pi i u_{mn}}-1}([e_{ij}\otimes e_{kl}, \eta^{a+b}_1]-
[e_{ij}\otimes e_{kl}, \eta^a_1])
\nonumber\\
&-\sum_{m,n=1}^N \frac{\al_{im}(\al^{-1})_{ml}\al_{kn}(\al^{-1})_{nj}}
{q^a\, e^{-2\pi i u_{mn}}-1}([e_{kl}\otimes e_{ij}, \eta^{b+a}_2]-
[e_{kl}\otimes e_{ij}, \eta^b_2])\,,
\label{DBLL}
\end{align}
where $u_{mn}=u_m-u_n$.

Finally, to calculate the Dirac bracket between the fields
$\eta(z)$ and $\eta(z')$ we must to perform the
sum
\begin{equation}
\{\eta_1(z),\eta_2(z') \}_{DB}
=\sum_{a,b\in {\bf Z} }\{\eta^a_1,\eta^b_2\}_{DB}z^a (z')^b.
\label{summ}
\end{equation}
Such sums require some care because separately
certain series in the expression for the Dirac bracket
(\ref{PBLL},\ref{DBLL}) do not converge. These must be
accurately combined in order to get a finite answer.

For example, if we choose $|q|<|z'|<|z|<1$, hence
$|q|<|z'/z|<1 $ and $1<|z/z'|<|q^{-1}|$, then the
second sum in
(\ref{DBLL}) gives a divergent series in (\ref{summ}). 
However if we add to the above terms the expression (\ref{PBLL}) 
we obtain a convergent series
and the desired $r$-matrix for the 
Lax matrix (\ref{Lax}) takes the following form

$$r(z,z')= \al_1\al_2[(E(z'/z)-E(z'))\sum_i e_{ii}\otimes e_{ii}+$$
\begin{equation}
+\sum_{i\neq j} (\phi(z'/z,u_{ji})-\phi(z',u_{ji}))e_{ij}\otimes e_{ji}]
\al^{-1}_1\al^{-1}_2\,,
\label{RMAT}
\end{equation}
where
$$\al_1=\al\otimes I\,, \qquad \al_2=I\otimes \al$$ 
and the function $E(z)$ can be represented in the form of the 
following series

\begin{equation}
E(z)=\sum_{a\in{\bf Z}\,a\neq 0} \frac{z^a}
{q^a -1}-1, \qquad |q|<|z|<1.
\label{SerE2'}
\end{equation}

~\\
{\bf Remark 1.} The classical $r$-matrices for the spin
Calogero-Moser systems were originally found in the 
paper \cite{ER}. In contrast to \cite{ER}  where the final formulae,
both for the Lax matrix and for the $r$-matrix, were
defined on the elliptic curve only after an additional auxiliary reduction,
our formulae (\ref{Lax}) and (\ref{RMAT}) are defined on the 
elliptic curve from the outset.

~\\
{\bf Remark 2.} The elliptic Calogero-Moser 
system with maximal spin sector can be also described 
in terms of a Lax matrix and $r$-matrix which are functions, 
rather than sections of some bundles on the elliptic curve.  
Namely, the Lax matrix (\ref{Lax}) and the 
$r$-matrix (\ref{RMAT}) are gauge equivalent
to the following matrix-valued functions on 
$\Si_{\tau}$
\begin{equation}
\begin{array}{c}
\displaystyle
l^{K}_{ij}(w)=\sum_{k,l=1}^N \al_{ik} B_{kl}(w)(\al^{-1})_{lj}, \qquad 
B_{ii}=-\frac{v_i}{2\pi i}, \\[0.5cm]
\displaystyle
B_{ij}(w)=-\sum_{k=1}^N \beta_{ik} \al_{kj}~
\frac{\te_{11}(w-u_j)\te_{11}(w+u_j-u_i)\te_{11}(u_i)\te_{11}'(0)}
{\te_{11}(w)\te_{11}(w-u_i)\te_{11}(u_i-u_j)\te_{11}(u_j)},\quad i\neq j,
\end{array}
\label{tL}
\end{equation}

\begin{equation}
\begin{split}
\tilde r(w,w')&=\sum_{i,j=1}^N (E(w-w')
+E(w')) e_{ij}\otimes e_{ji}- \\
&\qquad -\sum_{i,j,k,l=1}^N \al_{il}(\al^{-1})_{lk} 
(E(w-u_l) + E(u_l)) e_{ij}\otimes e_{jk},
\label{onA}
\end{split}
\end{equation}
where 
$$
w=\frac1{2\pi i}\ln z\,, \qquad w'=\frac1{2\pi i}\ln z'\,,
$$
and 
$$
\te_{11}(w)=\sum_{m\in{\bf Z}}
{\rm exp}\,(\pi i \tau (m+1/2)^2 + 2\pi i (m+1/2)(w+1/2)).
$$
The Lax matrix for the spin Calogero-Moser 
system in the form (\ref{tL}) was originally obtained 
in the paper by Krichever \cite{IK}, while 
the formula (\ref{onA}) for the corresponding $r$-matrix has been
given recently in \cite{VAD}.      

~\\
{\bf Remark 3.} In obtaining the result (\ref{RMAT}) we have only
used the definition of the Dirac bracket and {\it not}
the elliptic function identities which are
unavoidable for checking that (\ref{RMAT}) is indeed 
the $r$-matrix for the Lax matrix (\ref{Lax}).
Of course the functions our procedure yields as series are, as shown in the
Appendix, expressible as elliptic functions.

\vskip0.2in

The resulting $r$-matrix (\ref{RMAT})
is universal in a sense that it allows us to 
derive a classical $r$-matrix for the 
respective Hitchin system on the elliptic
curve with an arbitrary quasi-parabolic structure 
at the marked point. We will conclude this section by
elaborating upon this point.

Let $\cP$ be a parabolic subgroup of $SL_N(\bC)$. Then 
the Hitchin system on the elliptic curve with 
the corresponding  quasi-parabolic structure 
at the marked point is obtained from the 
maximal spin system described above upon the further
Hamiltonian reduction with respect to the 
following action of the parabolic subgroup $\cP$,
\begin{equation}
\al_{ij} \mapsto \sum_{k=1}^N q_{ik}\al_{kj}\,, \quad
\al_{ij} \mapsto \sum_{k=1}^N \beta_{ik} (q^{-1})_{kj}\,, \quad
u_i \mapsto u_i\,, \quad
v_i \mapsto v_i.
\label{Parabol}
\end{equation}
Here $q=(q_{ij})\in \cP$. The momentum map of 
the action (\ref{Parabol}) is found to be 
\begin{equation}
\mu\sp{P}\, : \, T^*{\cal M} \mapsto {\rm Lie}^*(\cP)\,, \qquad 
\mu\sp{P}[\al,\beta](X)=\tr (\al\beta X)\,, \quad X\in {\rm Lie}(\cP)\subset gl_N(\bC).
\label{mParabol}
\end{equation}
We note that the Lax matrix (\ref{Lax}) transforms under the 
action of $q\in \cP$ as 
\begin{equation}
l(z)\mapsto l^q(z)=q l(z) q^{-1},  
\label{transss}
\end{equation}
and this agrees with our general 
assumption (\ref{trans}). 

This stage of Hamiltonian reduction is finite dimensional and,
whatever parabolic subgroup $\cP$ and gauge fixing conditions are chosen,
our formula (\ref{viaDirac}) yields the desired $r$-matrix.
To proceed further with this final
Hamiltonian reduction we must now choose a concrete 
parabolic subgroup $\cP$ and impose the
constraints $\mu\sp{P}(\al,\beta)=0$ together with some gauge fixing conditions.
One could implement this simply using the Dirac reduction procedure 
once more to obtain finally an $r$-matrix for the elliptic spin Calogero-Moser 
system with the chosen quasi-parabolic structure.
An alternative method however exists, and we will use this. 
This alternate route proceeds by constructing a gauge invariant extension 
of the Lax matrix. What is this \emph{gauge invariant extension}?
We have said that our Lax matrix transforms as (\ref{transss}). A
gauge invariant extension of this is constructed by conjugating by a
compensating gauge transformation that ``undoes" this gauge transformation.
We imposed some gauge-fixing constraints $\chi\sp{a}=0$ so as to choose
(locally) one representative from each gauge orbit. Thus points in our
total space can be described (locally) by coordinates $(p,\tq)$, where
$p$ is a point on the constraint surface $\chi\sp{a}=0$ and $\tq$ describes
the gauge transformation along the orbit. The gauge invariant extension $l\sp P$ of
the Lax $l$ is
$$
l\sp{P}(z)= q\sp{-1} l(z)\, q$$
where if $l[p]$ transforms to $\tq\sp{-1} l\, \tq$ at the point $(p, \tq)$ then
$q(p,\tq)=\tq\sp{-1}$ so that $l\sp P (p,\tq)= l(p)$.
Therefore the twisted Lax matrix $l\sp P$ is a gauge invariant extension.
Now we have described in (\ref{rgauget}) how the $r$-matrix transforms under
a gauge transformation, the desired  $r$-matrix for the considered
Hitchin system with the quasi-parabolic structure $\cP$ are found 
to be 

\begin{align}
r^{P}(z,z')&=(q_1^{-1}q_2^{-1}r(z,z')q_1 q_2\nonumber \\
&\quad+ q_1^{-1}q_2^{-1}\{l_2(z') ,q_1\} q_2+
\frac12 [ q_1^{-1}q_2^{-1} \{q_1 ,q_2\},
q_2^{-1} l_2(z') q_2 ])\Big|_{on~shell}.
\label{RGadin}
\end{align}
Observing that $q(p)=Id$, this expression reduces to (\ref{viaDirac}),
and so explains the origin of the remark we made in deriving our theorem.
Thus by calculating (\ref{RGadin}) via the gauge invariant extension
we will obtain the same $r$-matrix as determined by Dirac reduction.
Reduction to the spin zero Calogero-Moser models yields the r-matrices
of \cite{Skrm, BSa, BSb}.

\section{Example 2: The Feigin-Odesskii Bracket Via Poisson Reduction}
In a recent paper \cite{O} the classical Sklyanin algebra
has been derived by Hamiltonian reduction to the moduli space of complex
structures on a principal $GL_2$-bundle of
degree one over an elliptic curve. In this section
we generalise the construction of \cite{O} to the
case of $GL_N$ in order to obtain
the classical Feigin-Odesskii algebra \cite{FO}.

Let $\Si_{\tau}$ be an elliptic curve and let $E_N$ be a principal
$GL_N$-bundle over $\Si_{\tau}$ defined by the following
gluing rules for a section $s$:
\begin{equation}
s(z+1)=I_1 s(z), \qquad
s(z+\tau)= \La(z) s(z).
\label{EN}
\end{equation}
Here
$$
\La(z)=I_2 \exp\left(-\frac{2\pi i z}{N}\right),
$$
$$
I_1=\left( \begin{array}{ccccc}
1 & 0 & \ldots & 0 & 0\\
0 & \ve & \ldots & 0 & 0 \\
~ & \ldots & \ldots & \ldots &~ \\
~ & \ldots & \ldots & \ldots &~ \\
0 & 0 & \ldots &0 &  \ve^{N-1}
\end{array}
\right),\qquad
I_2=\left( \begin{array}{ccccc}
0 & 1 & 0 & \ldots & 0 \\
0 & 0 & 1 & \ldots & 0 \\
\ldots & \ldots & \ldots &\ldots & \ldots \\
0 & 0 & \ldots & 0  & 1 \\
1 & 0 & \ldots & 0 &  0
\end{array}
\right),
$$
and $\ve=\exp\left(\frac{2\pi i}N\right)$.

The \v{C}ech cocycle condition determining this bundle follows from the
commutation relation
$$
I_1 I_2 = \ve^{-1} I_2 I_1.
$$
It easy to check that the holomorphic section
of the corresponding determinant bundle det($E_N$)
is just a $\te$-function with a single simple zero
on $\Si_{\tau}$. We therefore conclude that deg$(E_N)=1$.
Let us also note that
\begin{equation}
I_a=I^{a_1}_1 I^{a_2}_2 \qquad
a=(a_1,a_2)\in {\bf Z}_N\times {\bf Z}_N
\label{basis}
\end{equation}
forms a basis of $gl_N$.

In our construction of the Feigin-Odesskii bracket we
use deformations of the complex structure on $E_N$
that preserve the determinant bundle det($E_N$) of $E_N$.
In contrast with previous Section we will now use the Dolbeault picture
to describe the complex structures on $E_N$. The operator
\begin{equation}
d_{\bar A} =k\db + \bar A ~:~ \cA^{p,q}(\Si_{\tau}, E_N)
\mapsto \cA^{p,q+1}(\Si_{\tau}, E_N)\,, \qquad tr \bA(z,\bz)=0
\label{dbar}
\end{equation}
acts on the sections of $E_N$ and defines a complex structure
on $E_N$.
Here $\bA=\bA(z,\bz)d\bz$ is a $(0,1)$-connection of $E_N$:
\begin{equation}
\bA(z+1)=I_1 \bA(z) I^{-1}_1\,, \qquad
\bA(z+\tau)=I_2 \bA(z) I^{-1}_2.
\label{barA}
\end{equation}
The constant $k$ appearing in (\ref{dbar}) may be further identified with
a central charge.
Note that the traceless condition
$tr\bA(z,\bz)=0$ guarantees that the determinant of the deformed
holomorphic bundle $\tilde E_N$, defined by $k\db + \bar A$ coincides
with that of $E_N$.

Two complex structures $\bA$ and $\bA^f$ are called equivalent
if they are related by the following transformation
\begin{equation}
\bA \mapsto \bA^{f}=f^{-1}\bA f + f^{-1}k\db f
\label{gtran}
\end{equation}
where $f=f(z,\bz)$ is a smooth $SL_N$-valued function
on $\Si_{\tau}$ satisfying the
following quasi-periodicity conditions
\begin{equation}
f(z+1)=I_1 f(z) I^{-1}_1, \qquad
f(z+\tau)=I_2 f(z) I^{-1}_2.
\label{param}
\end{equation}

Now we define the total Poisson space $\bP$ that will be reduced
as a principal affinization over
a cotangent bundle to the space of smooth sections of $Aut(E_N)$.
A point on the respective base is identified with
a $GL_N$-valued field $g=g(z,\bz)$, satisfying the following
quasi-periodicity conditions
\begin{equation}
g(z+1)=I_1 g(z) I^{-1}_1, \qquad
g(z+\tau)=I_2 g(z) I^{-1}_2;
\label{supplf}
\end{equation}
and a point on the fibre of $\bP$ is determined by a
$(1,0)$-connection $\bA$ (\ref{barA}).

The Poisson brackets on $\bP$ are defined in the following way:
\begin{align}
\{\bA_{ij}(z,\bar{z}) ,\bA_{kl}(w,\bar{w})
\}&=(\bA_{il}(z,\bar{z})\de_{kj}-\bA_{kj}(z,\bar{z})\de_{il})\,\de(z-w)
\,\de(\bar{z}-\bar{w})\nonumber \\
&\qquad +
k(\de_{il}\de_{kj}-\frac{1}{N}\de_{ij}\de_{kl})\db\,\de(z-w)
\,\de(\bar{z}-\bar{w}),\nonumber \\
\{g_{ij}(z,\bar{z}) ,\bA_{kl}(w,\bar{w})\}&=(g_{il}(z)\de_{kj}-
\frac1N g_{ij}(z)\de_{kl})\,\de(z-w)
\,\de(\bar{z}-\bar{w}),
\label{RPB}\\
\{g_{ij}(z,\bar{z}) ,g_{kl}(w,\bar{w})\}&=0,\nonumber
\end{align}
where $k$ is a central charge.
The space $\bP$ with this Poisson structure is a particular
case of a general construction proposed by Polishchuk \cite{Po}.
If we supplement the transformations of $\bA$ (\ref{gtran}) with the
simultaneous transformation of the field $g$
\begin{equation}
\bA \mapsto \bA^{f}=f^{-1}\bA f + f^{-1}k\db f
\qquad
g \mapsto g^{f}=f^{-1}g f
\label{gtrans}
\end{equation}
then the equations (\ref{gtrans}) define a Poisson action
on $\bP$.

Let us consider the Poisson reduction with respect to the action
(\ref{gtrans}).
First we note that a generic field $\bA$ is in fact a pure
gauge\footnote{In other words this means that
there are no moduli of (semi)stable complex structures on $E_N$.}
\begin{equation}
\bA=f^{-1}k\db f
\label{fixing}
\end{equation}
and we can choose $\bA=0$ as an appropriate gauge
condition.
Further, there are no residual gauge symmetries. To see this,
consider a holomorphic function $f:{\bf C}\mapsto SL_N$ satisfying
the quasi-periodicity conditions (\ref{param}). By expanding in the basis
(\ref{basis}) we find that
$$f_a(z+1)=\ve\sp{-a_2}\, f_a(z),\qquad
f_a(z+\tau)=\ve\sp{a_1}\, f_a(z).
$$
Such a function must be a constant $f(z)=Id$.
Thus the desired Poisson quotient is parameterised by a
smooth section $L=L(z,\bz)$ of $Aut(E_N)$
\begin{equation}
L(z,\bz)=f(z,\bz)g(z,\bz)f^{-1}(z,\bz).
\label{Lax1}
\end{equation}

In order to calculate Poisson brackets on the reduced space
we have to construct a gauge invariant functional $L=L[g,\bA]$
which coincides with $L(z,\bz)$ on the surface of gauge fixing
$\bA=0$. For this we solve the equation (\ref{fixing}) for
$f(z,\bz)$ and substitute the solution into (\ref{Lax1}).
Thus we get the desired gauge invariant extension
\begin{equation}
L[g,\bA](z,\bz)=f[\bA](z,\bz)g(z,\bz)f^{-1}[\bA](z,\bz).
\label{invLax}
\end{equation}

The calculation of the Poisson bracket between the entries of the matrix
(\ref{Lax1}) is reduced to the calculation of an on-shell
expression using the Poisson bracket (\ref{RPB}) between the entries of
(\ref{invLax}).  Note, that in doing these calculations we do not
need to know the explicit dependence of $f_{ij}[\bA]$ on $\bA$.
The only terms that enter the on-shell expression are
\begin{equation}
r_{ij\,kl}(z,\bar{z};w,\bar{w})=
k\frac{\de f_{ij}(z,\bar{z})}{\de \bA_{lk}(w,\bar{w})}\Bigg|_{\bA=0}=
-k\frac{\de (f^{-1})_{ij}(z,\bar{z})}{\de \bA_{lk}(w,\bar{w})}\Bigg|_{\bA=0}.
\label{Green}
\end{equation}

Due to equation (\ref{fixing}) the function
$r_{ij\,kl}(z,\bar{z};w,\bar{w})$  turns out to
be a Green's function for the $\db$-operator
\begin{equation}
\db_z r_{ij\,kl}(z,\bar{z};w,\bar{w})=(\de_{il}\de_{kj}-\frac{1}{N}\de_{ij}\de_{kl})
\,\de(z-w)\,\de(\bar{z}-\bar{w}).
\label{Green1}
\end{equation}

Note that the
function (\ref{Green})  can be found in the
form $r(z,w)=r(z-w)$ and ``unitarity''  in the sense that
\begin{equation}
r(z)=-r_{21}(-z).
\label{unitar}
\end{equation}
In virtue of equations (\ref{barA}), (\ref{param}), we have that
$r(z)$ also possesses the following quasi-periodicity conditions
\begin{equation}
r(z+1)=(I_1\otimes Id)\, r(z)\, (I^{-1}_1\otimes Id), \qquad
r(z+\tau)=(I_2\otimes Id)\, r(z)\, (I^{-1}_2\otimes Id).
\label{period}
\end{equation}
Thus $r_{ij\,kl}(z)$ defines a
meromorphic function $r(z)\, :\,{\bf C}\mapsto sl_N \otimes sl_N $
with at most simple poles only at the points
$z=n+m\tau\,~n,m\in {\bf Z}$. The residue $t$ at the point $z=0$
is the Killing form of the Lie algebra $sl_N$
\begin{equation}
t=\sum_{i,j}(e_{ij}\otimes e_{ji}-\frac1N e_{ii}\otimes e_{jj}).
\label{tform}
\end{equation}
Following the paper \cite{BD} there is a unique meromorphic function $r(z)$
satisfying the above conditions and, in particular, it is a solution
of a classical Yang-Baxter equation
\begin{equation}
[r_{12}(z_1-z_2),r_{23}(z_2-z_3)]+[r_{12}(z_1-z_2),r_{13}(z_1-z_3)]+
[r_{13}(z_1-z_3),r_{23}(z_2-z_3)]=0.
\label{CYB}
\end{equation}

Returning then to the calculation of the
on-shell expression for the Poisson bracket
$$
\{L_1[g,\bA](z,\bar{z}), L_2[g,\bA](w,\bar{w})\}|_{\bA=0}
$$
there are two types of contribution.
The first type of contribution originates from the
Poisson bracket between $g(z)$ and $f[\bA](w)$,
and the respective expression is
\begin{align}
B_1&=2k^{-1}L_1(z,\bar{z})L_2(w,\bar{w})r(z-w)-
k^{-1}L_1(z,\bar{z})r(z-w)L_2(w,\bar{w})\nonumber \\
&\qquad - k^{-1}L_2(w,\bar{w})r(z-w)L_1(z,\bar{z}).
\label{Exp1}
\end{align}
The second set of terms arise from
the Poisson bracket between $f[\bA](z)$ and $f[\bA](w)$.
These have the following form
\begin{align}
B_2&=-k^{-1}r(z-w)L_1(z,\bar{z})L_2(w,\bar{w})+
k^{-1}L_1(z,\bar{z})r(z-w)L_2(w,\bar{w})\nonumber \\
&\qquad +k^{-1}L_2(w,\bar{w})r(z-w)L_1(z,\bar{z})-
k^{-1}L_1(z,\bar{z})L_2(w,\bar{w})r(z-w).
\label{Exp2}
\end{align}
Combining the expressions (\ref{Exp1}) and  (\ref{Exp2})
we obtain the classical quadratic $r$-matrix
\begin{equation}
\{L_1(z,\bar{z}),L_2(w,\bar{w})\}=\frac1k[L(z,\bar{z})\otimes L(w,\bar{w}),r(z-w)]
\label{LL}
\end{equation}
The Jacobi identity for the
Poisson bracket (\ref{LL}) follows from
the classical Yang-Baxter equation (\ref{CYB}).

In order to get
an explicit expression for the $r$-matrix (\ref{Green})
we use the basis (\ref{basis})
and certain automorphic functions on ${\bf C}$ satisfying
the following quasi-periodicity properties
\begin{equation}
\vf_a(z+1)=\ve^{-a_2}\vf_a(z) ,\qquad
\vf_a(z+\tau)=\ve^{a_1}\vf_a(z).
\label{vf}
\end{equation}
It is easy to see that
up to an overall factor for each $a\neq 0$
there exists a single
function $\vf_a$ satisfying
the above periodicity conditions (\ref{vf}) and having
at most simple poles only at the points
$n+m\tau,~n,m \in{\bf Z}$. For $a=(0,0)$ we simply set
$\vf_{(0,0)}(z)\equiv 1$.
Then the elliptic $r$-matrix takes
the following form
\begin{equation}
r(z)=\frac1N
\sum_{a\in {\bf Z}_N\times {\bf Z}_N~a\neq 0 }
\ve^{a_1 a_2}\vf_a(z) I_a\times I_{-a}.
\label{elli}
\end{equation}

Let us return to the Poisson bracket (\ref{LL}).
Consider the following set of functionals
\begin{equation}
C_{ij}=\int_{\Si_{\tau}} d^2z~ \psi(z,\bz) \db L_{ij}(z,\bz)
\quad i,j=1,\ldots N,
\label{Cas}
\end{equation}
for all smooth functions $\psi$ vanishing at the points
$n+m\tau,~n,m \in{\bf Z}$.
We will prove that this is a set of weak Casimir
functions\footnote{Functions $C_A$ are said to form
a set of weak Casimir functions if
they generate a proper ideal in the
Lie algebra of smooth functions with respect to the
Poisson bracket.} for the Poisson bracket (\ref{LL}).

To this end let $C$ be the matrix with entries (\ref{Cas}).
Straightforward calculation then shows that
\begin{align}
\{ C\otimes 1, L_2(w,\bar{w})\}&=
\int_{\Si_{\tau}} d^2z~ \psi(z,\bz) \db
\{ L_1(z,\bz), L_2(w,\bar{w})\} \nonumber \\
&=\int_{\Si_{\tau}} d^2z~ \psi(z,\bz)
\db\frac1k[L(z,\bz)\otimes L(w,\bar{w}),r(z-w)].
\label{calc}
\end{align}
Up to terms vanishing on the surface $C_{ij}=0$ (for all $i$, $j$)
the expression (\ref{calc}) takes the form
\begin{equation}
\{ C\otimes 1 , L_2(w,\bar{w})\}=\int_{\Si_{\tau}} d^2z~ \psi(z,\bz)
\frac1k[L(z,\bz)\otimes L(z,\bz), t\de(z-w)\,\de(\bz-\bw)]
\label{calcu}
\end{equation}
where $t$ is the residue of the r-matrix
$r(z)$ at the point $z=0$. Since $t$ is
adjoint invariant we see $\{ C\otimes 1 , L_2(w,\bar w)\}=0$.
A consequence of this result is that the zero level surface
for the functionals (\ref{Cas})
\begin{equation}
C_{ij}=0
\label{surf}
\end{equation}
is Poisson. That is, one can naturally define a Poisson bracket
on the surface (\ref{surf}) such that the embedding of the surface into the
initial space is a Poisson map.

Now the surface (\ref{surf}) is finite dimensional. In particular,
the matrix $L(z,\bz)$ satisfying the equations
(\ref{surf}) can be represented as
\begin{equation}
L(z)=\sum_{a\in {\bf Z}_N\times {\bf Z}_N }
S_a \vf_a(z) I_a,
\label{Laxxx}
\end{equation}
where $S_a$ are $c$-numbers parameterising the
surface (\ref{surf}).
Using some obvious properties of the automorphic functions $\vf_a$
one can easily show that the relation (\ref{LL})
defines a quadratic Poisson bracket between the coordinates
$S_a$. If we substitute the solution (\ref{Laxxx}) in
equation (\ref{LL})
this is just the desired Feigin-Odesskii bracket, which
is defined as a classical limit of the Feigin-Odesskii algebra
\cite{FO}.

Let us also note that all of the Casimir functions
of the Poisson brackets (\ref{LL}), (\ref{Laxxx})
can be constructed with the help of our proposed
reduction procedure.  To see this
we note the functional $\det g(z)$
yields a continuous set of Casimir functions
for the Poisson brackets (\ref{RPB}) of the unreduced space.
Because the functional $\det g(z)$ is also
gauge  invariant it defines
a continuous set of Casimir functions
$\det L(z)$ for the Feigin-Odesskii brackets
(\ref{LL}), (\ref{Laxxx}).
The desired algebraically independent
Casimir functions for (\ref{LL}), (\ref{Laxxx})
may then be defined as the following coefficients
of the Laurent expansion for $\det L(z)$
around the marked point
$z=0$\footnote{It is easy to see that
the residue of the function $\det L(z)$ at
the point $z=0$ is vanishing.}
\begin{equation}
{\cal C}_0=\oint_{\Gamma} \frac{\det L(z)}{z}, \qquad
{\cal C}_{-2}=\oint_{\Gamma} z\det L(z),
\quad\ldots\quad
{\cal C}_{-N}=\oint_{\Gamma} z^{N-1}\det L(z),
\label{Casimiry}
\end{equation}
where $\Gamma$ is a small contour around the
point $z=0$.

To conclude this section we would like to
mention a relation between the matrix (\ref{Laxxx}) and the Poisson
brackets (\ref{LL}), and the higher dimensional elliptic
top \cite{STS1}. The
equations of motion for this integrable system
are usually given in the form
\begin{equation}
\frac{d}{dt} L(z)=\frac12\{L(z),{\cal C}_{-2}\}\Big|_{l},
\label{eqofmo}
\end{equation}
where $\{ \cdot,\cdot \}\Big|_{l}$ stand for the
following linear Poisson brackets
\begin{equation}
\{L_1(z),L_2(w)\}\Big|_{l}=\frac1k([L_1(z),r(z-w)]-
[L_2(w),r_{21}(w-z)]),
\label{Llin}
\end{equation}
and ${\cal C}_{-2}$ is a quadratic Casimir function (\ref{Casimiry})
of the Poisson brackets (\ref{LL}), (\ref{Laxxx}).
The same equations of motion (\ref{eqofmo}) can however be
written in an alternative manner  with the aid of
the quadratic Poisson brackets (\ref{LL}):
\begin{equation}
\frac{d}{dt} L(z)=\{L(z), H\}
\label{eqofmo1}
\end{equation}
The Hamiltonian now has a very simple form
$$
H=-S_{0,0}.
$$
Note that the natural generalisations of the
elliptic tops \cite{STS1} associated to elliptic curves
with multiple marked points can be obtained analogously
in the framework of the our construction.
For this we must require the functions
$\psi(z,\bz)$, defining the weak
Casimir functions (\ref{Cas})
vanish at each marked point.

\section{Concluding Remarks}
In this paper we have shown how the technique of Hamiltonian reduction
enables us to calculate rather than guess some important 
ingredients in the theory of integrable systems.
Although the technique is not always rigorous in the field theory context
where conditionally convergent sums frequently arise,
this approach does give  us  some important geometric 
information about the objects in question.   

A stumbling block for the generalisation of our results 
to  curves of higher genus is the lack of  
a convenient basis both for meromorphic 
functions and for meromorphic 
differentials on a general Riemann surface. 
Thus in the paper \cite{VAD}, 
where a classical $r$-matrix is presented for 
Hitchin systems on an arbitrary Riemann surface of 
genus $g\ge 2$, it is mentioned that 
a formal expression for the $r$-matrix can 
be given in terms of a series in the 
Krichever-Novikov type basis \cite{KN}, \cite{KN1}. 
While we have shown that  sums of the type (\ref{summ})
can be performed for the elliptic curve, for an 
arbitrary curve of genus $g\ge 2$ the series yields only 
a small amount of information about the geometric nature of
the $r$-matrix in question.

{\bf Acknowledgements.} We are thankful to B.L. Feigin, A.M. Levin, A.V.
Odesskii, M.A. Semenov-Tian-Shansky and D. Talalaev for useful
discussions and comments. 
The work of AVZ, VAD and MAO is partially supported by the Grant
for Support of Scientific Schools 00-15-96557. The work of VAD
is supported by RFBR grant 00-02-17-956 and the grant INTAS 00-262;
the work of MAO and AVZ is supported by RFBR grant 00-02-16530;
the work of AVZ is also supported by the grant INTAS 00-561;
the work of HWB was supported in part by the NATO grant PST.CLG.976955.

\section*{Appendix: Representations of the Special Functions
$\phi(z,u,q)$ and $E(z,q)$.}

The meromorphic function $\phi(z,u)$ is defined by the
following properties. First, it is automorphic with respect to
the transformation $z \mapsto qz$, namely
\begin{equation}
\phi(q^{-1}z,u)=e^{2\pi i u}\phi(z,u).
\label{auto}
\end{equation}
Second, on the unit circle $|z|=1$ the function has one simple pole
at the point $z=1$ and the respective residue equals $1$.

It is easy to check that the function with the above properties
can be represented as
\begin{align}
\phi(z,u)&=\frac{1}{2\pi i}
\frac{\te_{11}(w+u)\,\te'_{11}(0)}{\te_{11}(w)\,\te_{11}(u)}
=\frac{1}{2\pi i}\frac{\sigma(w+u)}{\sigma(w)\,\sigma(u)}\,e\sp{-2\eta_1 u w},
\label{Phi}\\
&=\frac{1}{2\pi i}
\left( \frac{1}{\omega}+\zeta(u)-2\eta_1u
+\frac{\omega}{2}((\zeta(u)-2\eta_1u)^2-\wp(u))+\ldots \right).
\end{align}
Here
\begin{equation}
\te_{11}(w,\tau)=\sum_{m\in{\bf Z}}
{\rm exp}(\pi i \tau (m+1/2)^2 + 2\pi i (m+1/2)(w+1/2))
\label{TH}
\end{equation}
and $z=e^{ 2\pi i w}$. The function $\te_{11}$ is the unique odd theta
function,
$\sigma$,  $\zeta$ and $\wp$  are the Weierstrass functions
respectively of the same names, and $\eta_1=\zeta(\frac{1}{2})$.
Due to the automorphic properties of the $\te$-function (\ref{TH})
the right hand side of the equation
(\ref{Phi}) is in fact a function of $z$ and $q=e^{ 2\pi i\tau}$.

On the other hand, using the above properties one can find
series representations of the function (\ref{Phi})
and thus prove that the properties uniquely define the
desired meromorphic function.
Due to the poles of the function (\ref{Phi}) on each of the
circles
$|z|=|q|^{m},~m\in{\bf Z}$ the series representations depend on
the choice of the annulus. In deriving the $r$-matrix we use two
annuli, $1<|z|<|q^{-1}|$ and $|q|<|z|<1$. The series representations
of (\ref{Phi}) in these annuli are of the form
\begin{equation}
\phi(z,u)=\sum_{a\in{\bf Z}} \frac{z^a q^a e^{2\pi i  u } }
{q^a e^{2\pi i  u }-1}, \quad 1<|z|<|q^{-1}|\,, \quad u\neq
a+b\tau,~
a,b\in{\bf Z},
\label{Ser1}
\end{equation}
and
\begin{equation}
\phi(z,u)=\sum_{a\in{\bf Z}} \frac{z^a  }
{q^a e^{2\pi i  u }-1}, \quad |q|<|z|<1\,, \quad u\neq a+b\tau,~
a,b\in{\bf Z}.
\label{Ser2}
\end{equation}

The meromorphic function $E(z)$ is defined by analogous
properties. It is automorphic with respect to
the transformation $z \mapsto q^{-1}z$
\begin{equation}
E(q^{-1}z)=E(z)+1
\label{autoE}
\end{equation}
On the unit circle $|z|=1$ the function has one simple pole
at the point $z=1$ and the respective residue equals $1$.
Finally, the function $E(z)$ has a vanishing moment
in the annulus $1<|z|<|q^{-1}|$, namely
\begin{equation}
\oint_{|z|=|q|^{-1/2}} \frac{dz}{2\pi i z} E(z)=0.
\label{zerom}
\end{equation}

Using the $\te$-function (\ref{TH}) we may represent
$E(z)$ as
\begin{equation}
E(z,q)=\frac{1}{2\pi i}
\frac{\te'_{11}(w,\tau)}{\te_{11}(w,\tau)}-\frac12
=\frac{1}{2\pi i}\left(\zeta(w)-2\eta_1\,w\right)-\frac12.
\label{E}
\end{equation}

The series representations for $E(z)$
in the annuli $1<|z|<|q^{-1}|$ and $|q|<|z|<1$
are of the form
\begin{equation}
E(z)=\sum_{a\in{\bf Z}\,a\neq 0} \frac{z^a q^a  }
{q^a -1}, \qquad 1<|z|<|q^{-1}|,
\label{SerE1}
\end{equation}
and
\begin{equation}
E(z)=\sum_{a\in{\bf Z}\,a\neq 0} \frac{z^a  }
{q^a -1}-1, \qquad |q|<|z|<1.
\label{SerE2}
\end{equation}

\end{document}